\documentclass[letterpaper, 10pt, conference]{ieeeconf}  

\IEEEoverridecommandlockouts                              
\usepackage{romannum}
\usepackage{amsmath}
\usepackage{amssymb}
\usepackage{microtype}
\usepackage{graphicx}
\graphicspath{{./figures/}}
\usepackage{subfig}
\usepackage[margin=1.75cm, headheight=10pt]{geometry}
\makeatletter
\let\NAT@parse\undefined
\makeatother
\usepackage{hyperref}
\usepackage{listings}
\usepackage{textcomp}
\usepackage{gensymb}
\usepackage{mathtools}
\usepackage{atbegshi}    
\usepackage[skipbelow=\topskip, skipabove=\topskip]{mdframed}
\usepackage{multirow}
\usepackage{array}
\usepackage[font=small]{caption}
\newcolumntype{P}[1]{>{\centering\arraybackslash}p{#1}}
\usepackage{xcolor}
\title{\LARGE \bf Indoor Distance Estimation using LSTMs over WLAN Network}
\author{Pranav Sankhe, Saqib Azim, Sachin Goyal, Tanya Choudhary, \\ Kumar Appaiah, Sukumar Srikant \\
Indian Institute of Technology, Bombay, India \\
$\{ \text{pranav\_sankhe, saqib\_azim, sachin\_goyal, tanya.choudhary}\}$@iitb.ac.in \\
akumar@ee.iitb.ac.in \hspace{1cm}   srikant@sc.iitb.ac.in
\thanks{This work was supported by a grant from the Department of Science and Technology, Government of India (INSPIRE DST/INSPIRE/04/2014/001392). \newline Part of the work was supported by Bharti Centre for Communication in IIT Bombay \newline \textbf{978-1-7281-2082-9/19/\$31.00 \textcopyright2019 IEEE}}}
\begin{document}
\captionsetup[figure]{labelfont={bf}, labelformat={simple}, labelsep=colon, name={Figure}}
\maketitle
%
\begin{abstract}
The Global Navigation Satellite Systems (GNSS) like GPS suffer from accuracy degradation and are almost unavailable in indoor environments. Indoor positioning systems (IPS) based on WiFi signals have been gaining popularity. However, owing to the strong spatial and temporal variations of wireless communication channels in the indoor environment, the achieved accuracy of existing IPS is around several tens of centimeters. We present the detailed design and implementation of a self-adaptive WiFi-based indoor distance estimation system using LSTMs. The system is novel in its method of estimating with high accuracy the distance of an object by overcoming possible causes of channel variations and is self-adaptive to the changing environmental and surrounding conditions. The proposed design has been developed and physically realized over a WiFi network consisting of ESP8266 (NodeMCU) devices. The experiments were conducted in a real indoor environment while changing the surroundings in order to establish the adaptability of the system. We compare different architectures for this task based on LSTMs, CNNs, and fully connected networks (FCNs). We show that the LSTM based model performs better among all the above-mentioned architectures by achieving an accuracy of $\bf{5.85}$ cm with a confidence interval of $\bf{93\%}$ on the scale of ($\bf{8.46}$ m x $\bf{6.98}$ m). To the best of our knowledge, the proposed method outperforms other methods reported in the literature by a significant margin. \\ \textit{Index Terms} - Indoor Localization; WiFi; Received Signal Strength Indicator (RSSI); Long Short-Term Memory Network (LSTM).
\end{abstract}
\section{INTRODUCTION}
Indoor positioning systems are aimed at solving the problem of localization of objects and devices in closed rooms or buildings where GPS signals cannot reach due to high attenuation. The exact location of objects relative to the environment is a piece of crucial information for asset tracking, security, human-computer interface (HCI) applications as well as tasks like helping someone to find his or her way around an unknown building. \par
Currently, the Global Positioning System (GPS) is used for outdoor localization. GPS cannot be used indoors because the physics of radio propagation rules out the reception of weak GPS microwave signals indoors. Also, the reported accuracy of GPS is around $4$ m and hence is insufficient for high accuracy demanding indoor positioning. Consequently, a lot of work has been done to develop similar kind of systems for closed indoor environments with sub-meter range accuracy. \par
The existing indoor positioning systems are based on acquiring various signal parameters such as Received Signal Strength Indicator (RSSI), Channel State Information (CSI), Angle of Arrival (AoA), Time Difference of Arrival (TDoA) in case of multichannel communication, etc. These methods use techniques like trilateration and fingerprinting to localize the object. Recently deep learning has been successfully applied to solve a wide spectrum of challenging problems in the field of NLP, computer vision, medical image processing, etc. These models can extract complex features from the training data. CNN, RNN, and LSTM models have yielded high accuracy, especially in the classification tasks.\par 
Taking inspiration from the success of neural networks in classification tasks, we formulate the indoor localization problem as a classification task based on RSSI values acquired from the WiFi signals. We compare different architectures for this task based on LSTMs, CNNs, and fully connected networks. We show that LSTM based model performs better compared to all the architectures mentioned above since they have an inherent ability to learn spatial as well as temporal patterns making them suitable for indoor localization tasks. We discuss more over the evaluation of different ML architectures in Section \Romannum{3}. \par
To the best of our knowledge, LSTMs have not been used previously on RSSI values to solve the problem of indoor localization. One of the crucial contribution of this paper is to incorporate the time dependence of RSSI values and bring out the rationale behind why LSTMs are well suited to develop systems based on RSSI. The relation between the distance and the RSSI value depends on the topology of the indoor environment as well. To model the environmental topology, we propose a unique set-up consisting of 4 stationary WiFi nodes. We discuss the rationale behind the choice of four stationary nodes to model the multipath fading effects and one moving node (which is the object being tracked) to model the shadowing effects in greater detail in Section \Romannum{3}. Section \Romannum{4} discusses about the detailed experiments and the results. Finally in Section \Romannum{5} we talk about the possible future extensions of this work. \par
\section{LITERATURE REVIEW}
\subsection{Indoor Positioning Methods}
We give a brief description of various approaches adopted for indoor positioning problem. \par
\textbf{RSSI} based approaches are most common and easily deployable for indoor localization (\cite{sadowski_spachos_2018}, \cite{832252}, \cite{chintalapudi_iyer_padmanabhan_2010}, \cite{Ferris:2007:WUG:1625275.1625675}) since RSSI values can be acquired from any WiFi-based devices such as mobiles, laptops, etc. Sadowski et al. \cite{sadowski_spachos_2018} calculate the distance of the receiver from each of the transmitter using the path loss model based on RSSI values acquired by the receiver. The average reported accuracy in these approaches ranges from $2$ m - $4$ m. Distance estimation using the path loss model is based on the assumption that all the points lying on the boundary of a circle with the transmitter at the center will have the same RSSI which is not true due to asymmetry attributed by shadowing and multipath effects. \par
\textbf{Angle of Arrival Based}: CSI contains information about the channel between sender and receiver at the level of individual subcarriers for each pair of receiving and transmitting antennas \cite{Halperin:2011}. Hence AoA can be acquired using the CSI. Most of the techniques available in the literature that use AoA are either not deployable or not universal. ArrayTrack \cite{Xiong:2013} require $6$ - $8$ antennas, LTEye \cite{Kumar:2014LTE} require rotatory antenna and Ubicarse \cite{Kumar:2014} require motion sensors on the tracking device as well as demands the user to rotate the device by at least $180\degree$ thereby limiting their usage and scalability. \par
In \textbf{Fingerprinting} based approaches (\cite{Azizyan} -  \cite{Liu:2012:PLW:2348543.2348581}) the area of interest is scanned and a database of recorded signal characteristics is created. During test time, one tries to map signal features with the nearest point in database using multivariate analysis, support vector machines \cite{Halperin:2011} and KNN \cite{Wilkinson}. These approaches suffer from an inherent limitation of not being able to extract relevant and complex features from the data and use them as the deep learning based models. \par 
\textbf{Deep Learning Based Models}: Previously, approaches deploying FCN \cite{WangGao}, RNNs \cite{8267833} and CNNs \cite{CNN_Ibrahim}  for localizing object used the acquired RSSI, CSI or AoA values to train the model. Lukito et al.  \cite{8267833} perform classification of the object location using the signal RSSI and report a classification accuracy of $83\%$. But the locations are several tens of meters apart and hence does not account to centimeter level accuracy. Wang et al. \cite{WangGao} used 4 layer FCN to predict probabilistic locations based on CSI values. \par
\textbf{Other Methods}: There have been some attempts at using ultrasonic sensors (Chen et al. \cite{ChenGao}) for computing the time difference of arrival (TDoA) and localize the object thereafter. Huichao et al. \cite{HLv} use visible light communication and propose a differential detection based positioning algorithm to improve accuracy. Some attempts at using RFIDs (\cite{Wang:2013}, \cite{Wang:2013:DWM:2486001.2486029}, \cite{Wang:2014:RVT:2619239.2626330}) have also been made. Wang et al. \cite{WangYu} use smartphone\textquotesingle s magnetic and light sensors along with LSTM model. But none of the above approaches is scalable because of the extra hardware requirements compared to WiFi access point signal based methods like RSSI.
\subsection{Long Short-Term Memory (LSTM) Networks}
\begin{figure}[h!]%
    \centering
    \subfloat{{\includegraphics[width=8cm]{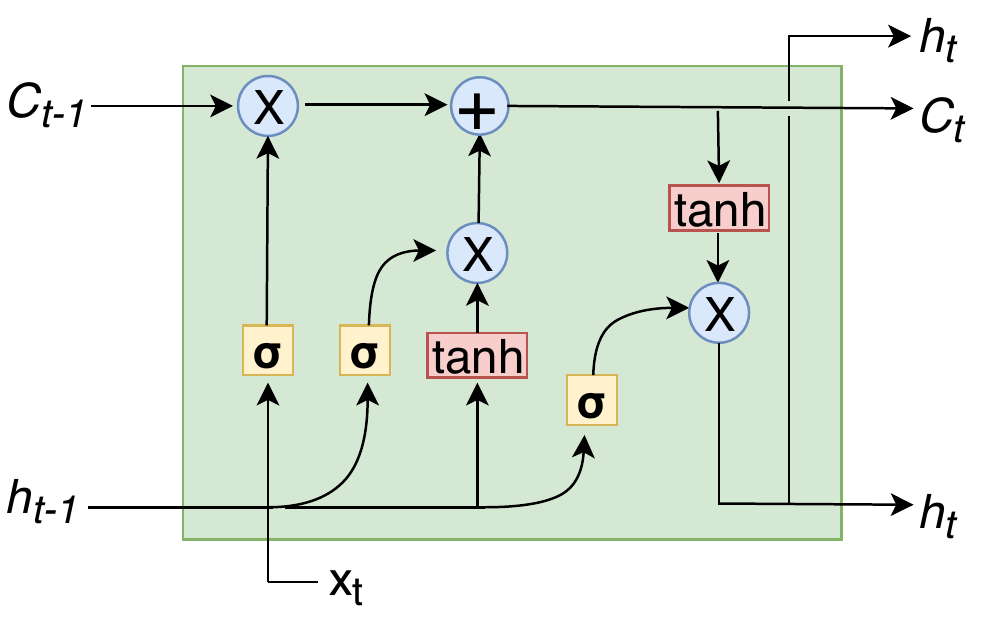}}}%
    \caption{A representation of LSTM Cell \cite{colah}. $C\textsubscript{t}$ represents cell state, $h\textsubscript{t}$ represents cell output and $\sigma$ represents sigmoid layer}%
    \label{fig:lstm_cell}%
\end{figure}
LSTMs are special kind of RNNs designed to remember and learn the time dependency of data. Here we briefly describe LSTM architecture \cite{colah}. Key to their structure is the \textit{cell state} to which information can be added by the cells through gates. Gates are sigmoid neural network layers which optionally let the information pass through and are of three types:
\begin{enumerate}
    \item \textit{Forget Gates}: New time step input is passed through a sigmoid layer and multiplied with current cell state to delete the information not required further.
    \item \textit{Input Gate Layer}: Input is passed through a tanh layer and selected values are added to the current cell state.
    \item \textit{Output Gate}: The current cell state is run through the tanh layer, and desired values(controlled using sigmoid) are output.
\end{enumerate}
\section{APPROACH}
\subsection{System Design}
\begin{figure}[h!]%
    \centering
    \subfloat{{\includegraphics[width=8cm]{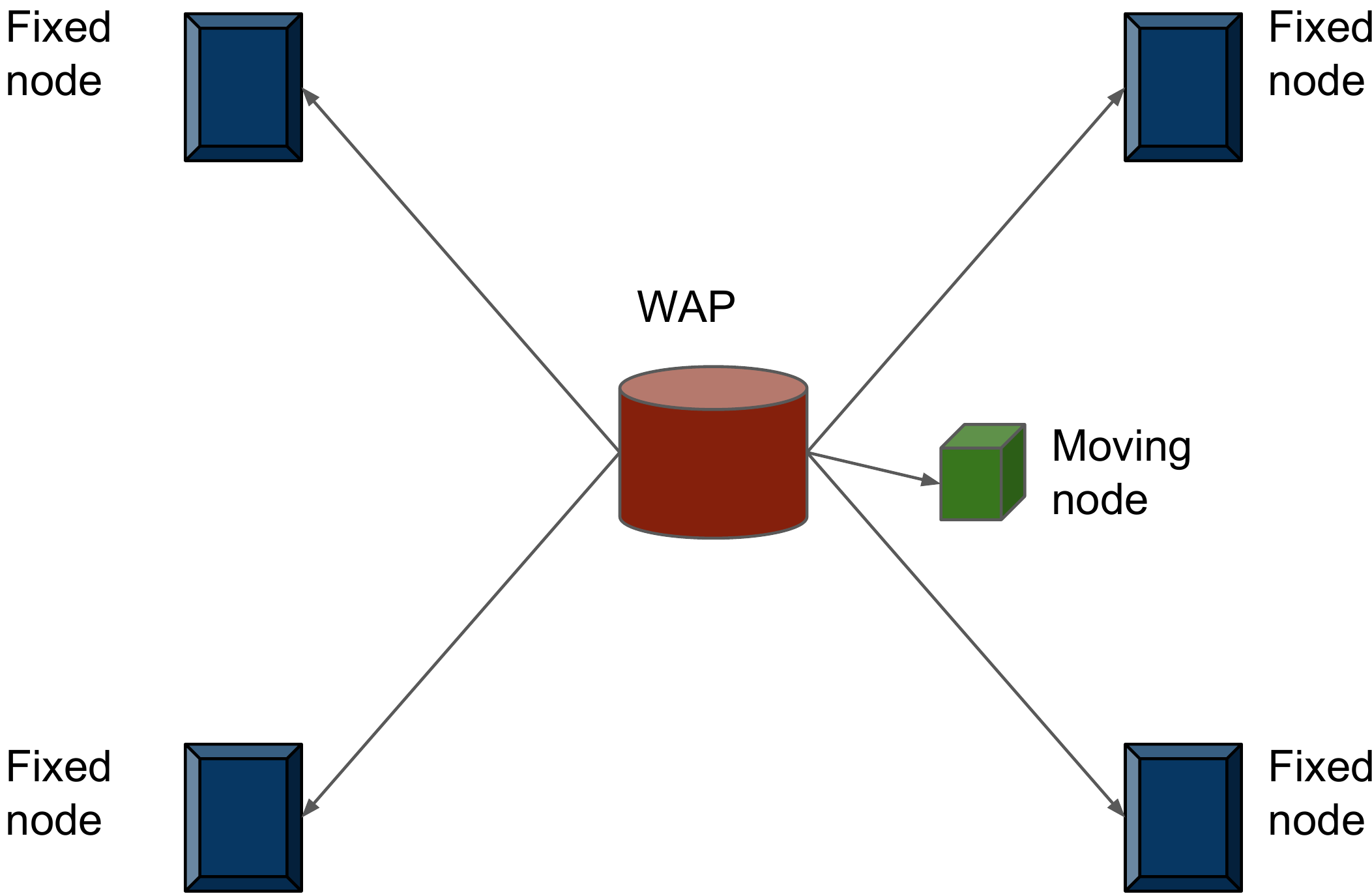}}}%
    \caption{WiFi Network Model consisting of a Wireless Access Point (also acting as reference), four fixed nodes at the corners of rectangular arena and a moving or target node whose distance from WAP needs to be estimated. The position of the fixed nodes with reference to WAP is known while the distance of target node from WAP needs to be estimated.}
    \label{fig:system}%
\end{figure}
As shown in Fig.\ref{fig:system}, our distance estimation system is a WiFi network consisting of a Wireless Access Point (WAP) and five WiFi clients (nodes). The WAP sets up the network to which the nodes can connect to. We classify these nodes into two distinct categories based on their position and motion: the fixed nodes and the moving or target node. The fixed nodes are placed at the corners of the indoor environment while the target node, which is to be tracked and whose distance from the WAP needs to be estimated, is allowed to move unobstructively inside the room. We have used four fixed nodes in our experiments. One can potentially use more number of nodes to ensure that the entire desired area is covered. In order to estimate the distance between the target node and the WAP, we use the dependence of signal strength received at the nodes on their distances from WAP and the topology of the indoor environment. This dependence is governed by path loss models like shadowing and multipath models, and the machine learning architecture has been designed to consider these path loss models for accurate distance estimation. \par
Shadowing is the effect which causes received signal power to fluctuate due to objects obstructing the propagation path between transmitter and receiver. Shadowing is a large-scale effect, as it corresponds to substantial deviation of the RF signal from its mean due to significant obstacles which create shadow zones causing deep fades if a receiver enters them. \par
From empirical measurements, it has been shown that the difference between the average and the actual path loss follow a log-normal distribution. Expressing the path loss in dB, we have
\begin{equation} \label{eq:power_path_loss_relation}
    P_L(d) \text{[dB]} = P_L(d_0) + 10\gamma \log \bigg( \frac{d}{d_0} \bigg) + X_{\sigma}
\end{equation}
where, $P_L(d)$ is the  power received at distance $d$, $P_L(d_0)$ is the  power received at the reference distance $d_0$,  $X_\sigma \sim \mathcal{N} ($0$,\sigma^2)$ describes the random shadowing effects and $\gamma$ is the path loss exponent. \par
A log-normal shadowing model is more suitable for the indoor localization problem as it provides several parameters which can be configured by machine learning system according to different indoor environments. The shadowing effect is spatial in nature \cite{puccinelli_haenggi_2006}. Therefore, the data required for learning the parameters of the shadowing model should have variation in distance of the target node from WAP. This information about the change in distance will be indicated by signal strength value measured at the target node, thus enabling the machine learning architecture to learn parameters of the shadowing model. \par
The shadowing model captures spatial variation only. But in wireless communication, the channel changes with time as well. In order to capture these temporal variations, we consider a multipath model, which is a combination of both the temporal and spatial model. Multipath effects are unwanted and can be mitigated. The unreliability of wireless networks can largely be attributed to multipath fading, and hence, it causes fairly large deviations in signal strength.
%
%
Multipath fading depends on the topology of the environment where the nodes are deployed \cite{puccinelli_haenggi_2006}. In wireless sensor networks, multipath fading is considered only for static nodes. Hence, in order to model the multipath effects and thus, the topology of the surroundings in the network, we place static WiFi nodes at the corners of the room. \par
\begin{figure}[h!]
    \centering
    \subfloat{{\includegraphics[width=8.5cm]{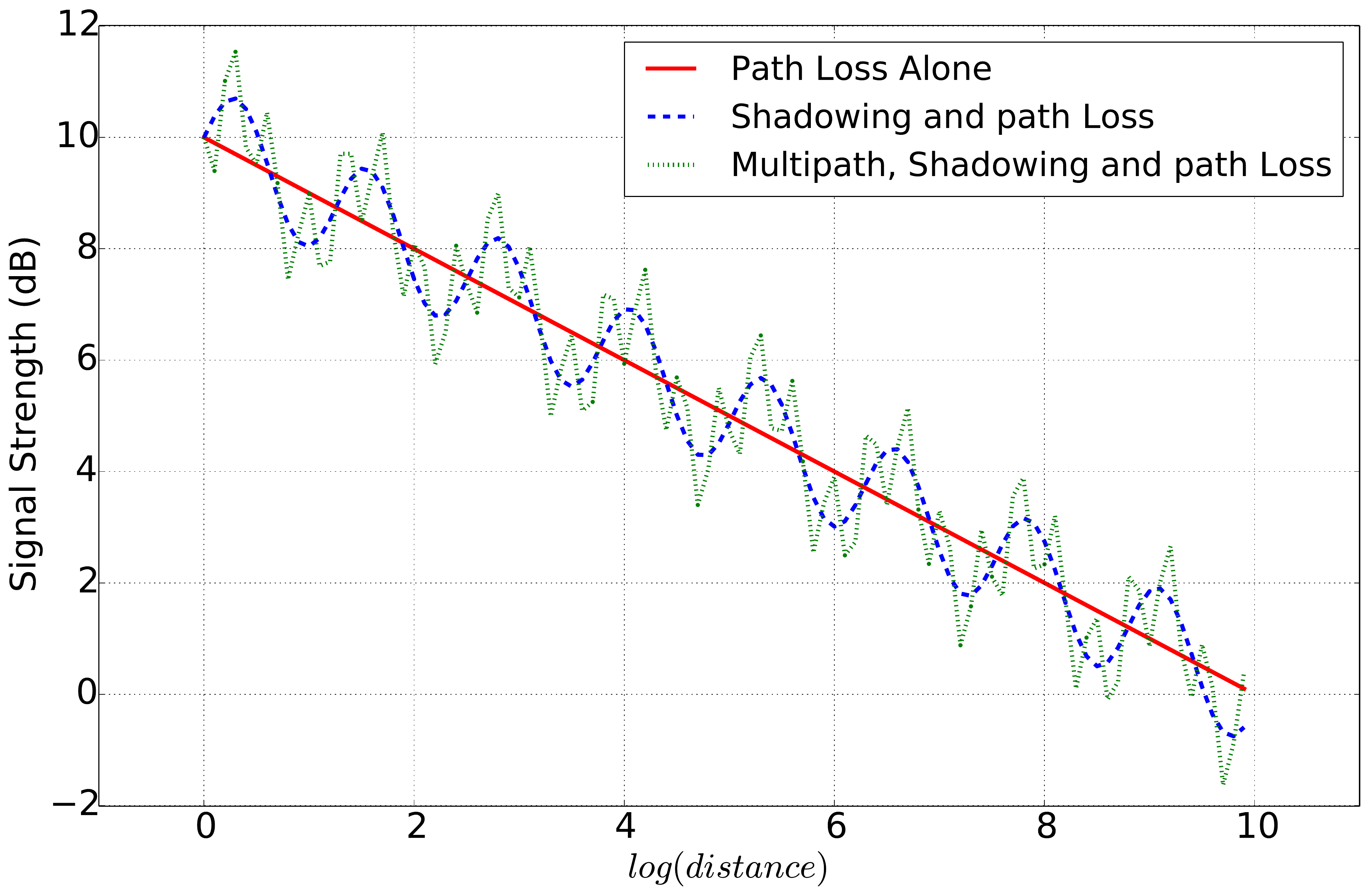}}}%
    \caption{This figure compares the variation of signal strength and distance when we consider the various propagation models in Wireless Networks. The dotted plot represents the signal strength variation when we consider all the path loss models and is the most accurate representation of the actual scenario amongst the three models described.}%
    \label{fig:signal_distance_plot}%
\end{figure}
Our machine learning architecture based on signal strength values received at the static WiFi nodes in the network learns the multipath fading model and hence, the topology of the indoor environment. Since the static nodes help to model the topology of the surroundings, our system is adaptable to any indoor environment which we demonstrate in our experiments by validating our trained model on four different datasets, all recorded with varying surrounding indoor conditions.
\subsection{Machine Learning Architecture}
We measure the signal strength at target node as it changes its position with time. It is assumed that the motion of target node follows a smooth trajectory. We measure the received signal strength at each position in the arena by varying the trajectory followed to reach there. We observed that in addition to the distance from the WAP, the trajectory followed by the target node to arrive at a particular position significantly affects the measured signal strength. This implies that the received signal strength at any time instant $t$ is correlated with its past values. We intend to exploit these temporal(time) correlations to predict the distance of the target node from the WAP. \par
We formulate our problem of predicting the distance of the target node from the WAP as a classification task by dividing the range of possible object distance values into $30$ bins of equal size. The object distance is then classified over these bins using deep learning methods. To this end, we introduce and compare three different ML architectures for this task, which are based on recurrent, convolutional, and fully connected neural network, respectively.
\newline
\subsubsection{Fully Connected Neural Network}
The fully connected neural network is the simplest type of artificial neural network where the information moves in the forward direction from the input layer, through the hidden layer and to the output layer. There are no cycles or loops in this kind of network. In our implementation, we used a fully connected architecture consisting of two hidden layers of size $64$ and $128$ neural units respectively and an output layer of size $30$ based on the number of distance classification bins. The input layer size depends on the dimension of the input data chosen, which is explained in the next section.
\newline
\subsubsection{Convolutional Neural Network}
CNNs have been successfully used in numerous machine learning based applications such as image classification, object detection, etc. These are very powerful in terms of capturing the spatial relations in the input data but lack the ability to learn any time-series dependence in the input. Our CNN architecture consisted of $5$ convolution layers with the first two layers having $8$ and $16$ filters each of size $3 \times 3$ respectively. This is followed by $3 \times 3$ max-pooling layer and dropout layer with dropout parameter ($\lambda=0.75$). The following $3$ convolution layers have $32$, $32$ and $64$ filters each of size $3 \times 3$.
%
\newline
\subsubsection{Long Short-term Memory Networks}
\begin{figure}[h!]%
    \centering
    \subfloat{{\includegraphics[height=5.5cm]{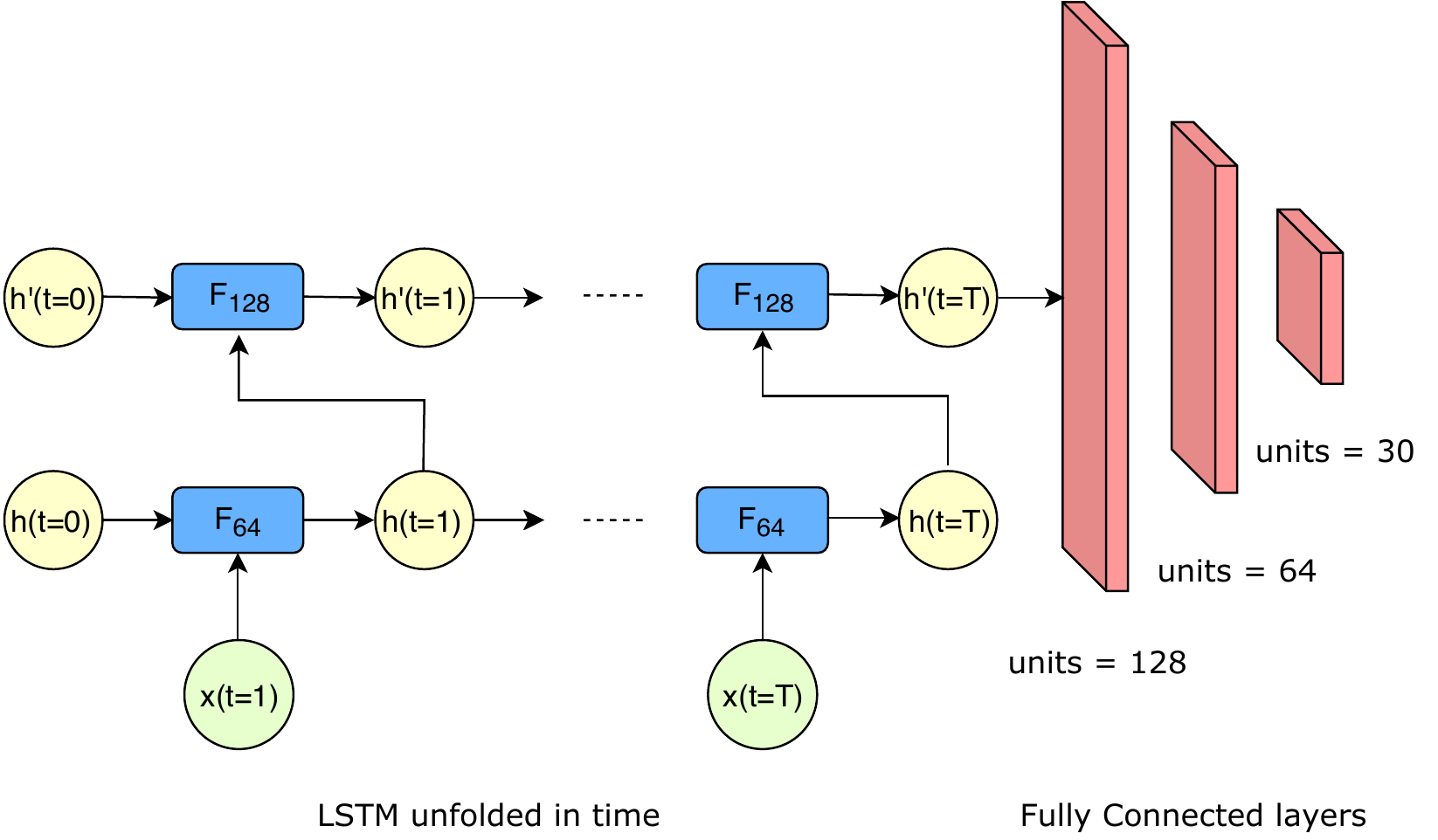}}}%
    \caption{Block diagram of LSTM based architecture unfolded in time. $2$ LSTM multirnn layers followed by fcn layers with relu activation function. $h(t)$ and $h^{'}(t)$ represent the cell states at time $t$, $x(t)$ represents input features at time $t$, while $F_{n}$ represents LSTM cell with cell state size as $n$}%
    \label{fig:ml_model}%
\end{figure}

We use LSTMs for time series modeling of received signal strength values as LSTMs have performed well, especially with the classification of time series sequences and text sequences (\cite{sentencestate}, \cite{Pengfei}). LSTMs are known to learn the long-term dependency as well as spatial and temporal patterns in the input data. Using LSTMs for time series modeling of RSSI values is one of the main reasons for the high accuracy achieved by our proposed model. As shown in Fig.\ref{fig:ml_model}, our LSTM based deep learning architecture consists of $2$ LSTM layers, each with $64$ and $128$ cell units respectively. The output of these LSTM layers is in $128$-dimensional feature vector space. Since the distance classification has to be done among $30$ classes, the output of LSTM layers is fed to a fully connected network (FCN) which gradually reduces the output from $128$-dimensional to $30$-dimensional feature space. The FCN is implemented with $2$ hidden layers having $64$ and $32$ neural units and rectified linear unit (ReLU) as non-linear activation function. Softmax function is applied to the output of last FCN layer to get the final classification scores over which cross entropy loss is used for training the model.
\subsection{Data Preprocessing}
This step involves cleaning the recorded distance data of the moving node from the WAP. To remove outliers in the data, a median filter of suitable window size is used, which helps to mitigate noise to a great extent and in turn allows better training of the ML architecture. Before feeding to the ML network, filtered data is appropriately shifted to zero-mean and divided by the standard deviation to normalize the range of independent variables.
\section{EXPERIMENTS \& RESULTS}
\subsection{Experimental Setup}
The entire experimental setup has been carried out on a horizontal 2D plane in a rectangular room (dimension $8.46$ m $\times$ $6.98$ m) consisting of various objects such as chairs, desks, etc. A central rectangular arena ($4.14$ m $\times$ $2.86$ m) inside the room was used for the localization setup. Four static reference nodes (NodeMCUs) were fixed at four corners of the arena, and a router (TP-Link TLWR840N) acting as a Wireless Access Point (WAP) was placed near the center of the arena. The object being tracked (whose distance from the WAP needs to be estimated) is placed inside the arena and is installed with the target node. All the nodes are wirelessly connected in a star network mode to a separate coordinator node, placed outside the arena. These five nodes ($4$ reference nodes and $1$ target node) send RSSI values from their connection with WAP to the coordinator node every $50$ ms. The task of the coordinator node is to receive RSSI values and send these to a Central Computation Unit (CCU). To collect data of RSSI values and corresponding distance of target node from WAP, we attach the target node to a bot programmed to move along a fixed path in the arena. To get the distance of the robot from WAP, a ceiling-mounted camera is used, with a frame rate of $20$ fps and localization accuracy within a range of $4$ mm. Each frame from the camera was processed to detect a circular red-patch placed on top of the robot and thus estimate its distance from the WAP. The distance estimated with the camera-based system served as the ground truth of the distance between the target node and WAP for our system. \par
The entire data for target object distance from WAP and corresponding RSSI values were collected on $4$ different days at different timings to ensure proper variation, if any, due to environmental changes. Each time, the experimental room setup was changed to create a sense of a different environment by varying the obstructions around the arena like chairs and table positions.
\subsection{Training and Evaluation}
The entire neural network architecture has been implemented and trained using TensorFlow. While the architecture of individual networks is different, they share certain similarities. All the networks are trained using the same input features and dimension. The input to all the architectures is a 3-dimensional matrix of size [$Bs$ $\times$ $W$ $\times$ $N$], where $Bs$, $W$ and $N$ represent training batch size, LSTM time steps and the total number of WiFi nodes respectively. $W$ acts as the context window in the CNN architecture and was empirically chosen to be $20$. Hence we reshape our data into matrices of the above dimensions. The network gradients were clipped to a maximum of $10$ as it led to better and stable training. The training of all networks has been done using Adam optimizer \cite{Kingma}. The network parameters are initialized using Xavier initialization \cite{glorot}, and cross-entropy loss function is used. All the networks perform classification over classes of possible object distances from the WAP. We train the model on NVIDIA GTX Titan GPU. It takes around $6$ hours for $3000$ training iterations. \par
We train and test our model separately on the $4$ different datasets collected from our lab to ensure the validity of our system under varying and different conditions. For completeness, we provide the list of all hyperparameter values for the LSTM architecture after final tuning in Table \ref{table:hyperparameter_values}.
\begin{table}[h!]
    \centering
    \begin{tabular}{ |c|c|c| } 
        \hline
        \textbf{Hyperparameters} & \textbf{Typical Values}  \\
        \hline
        Batch Size ($Bs$) & $1024$  \\ 
        LSTM Time Steps ($W$) & $20$ \\ 
        Maximum Gradient Norm & $10$ \\ 
        Learning Rate & $10^{-4}$ \\ 
        LSTM Sizes & $2$ layers $: 64$, $128$ cells \\
        Dropout parameter ($\lambda$) & $0.75$ \\
        \hline
    \end{tabular}
    \caption{List of all network Hyperparameter values}
    \label{table:hyperparameter_values}
\end{table}
%
%
\subsection{Results}
As stated earlier, in our experiments, object distance varies from $1.51$ cm to $3.41$ m. We formulate the localization problem as a classification task over $30$ equal bins each of length $l\textsubscript{bin}$ equal to $11.73$ cm. Our model predicts the label ($0$ - $29$) of the bin in which the object distance lies and reports the center of the predicted bin as the predicted distance. Two different accuracy metrics have been used to validate our proposed localization system:
\begin{enumerate}
    \item \textbf{Confidence Interval}: The center of the predicted bin is reported as the predicted distance. If the LSTM classifies correctly and hence predicts the correct bin, the upper bound on the error in the predicted distance is half of the bin length ($l\textsubscript{bin}$) which is $5.85$ cm. \par 
    As we can see in Fig.\ref{fig:accuracy_comparison}, the highest accuracy is obtained by the LSTM architecture. The classification accuracies achieved by LSTMs, CNNs and FCs are $93\%$, $65\%$ and $50\%$ respectively. The LSTMs achieve classification accuracies of $93.94\%$, $92.51\%$, $93.89\%$ and $92.99\%$ on the $4$ datasets respectively. Thus our model assures a $5.85$ cm upper bound on the error in the distance with the confidence intervals same as the classification accuracy. The accuracies are reported on the scale of ($8.46$ m $\times$ $6.98$ m). Fig.\ref{fig:accuracy_comparison} depicts the variation of classification accuracy as the three architectures get trained on dataset $1$.
    \begin{figure}[h!]%
        \centering
        \subfloat{{\includegraphics[width=8cm]{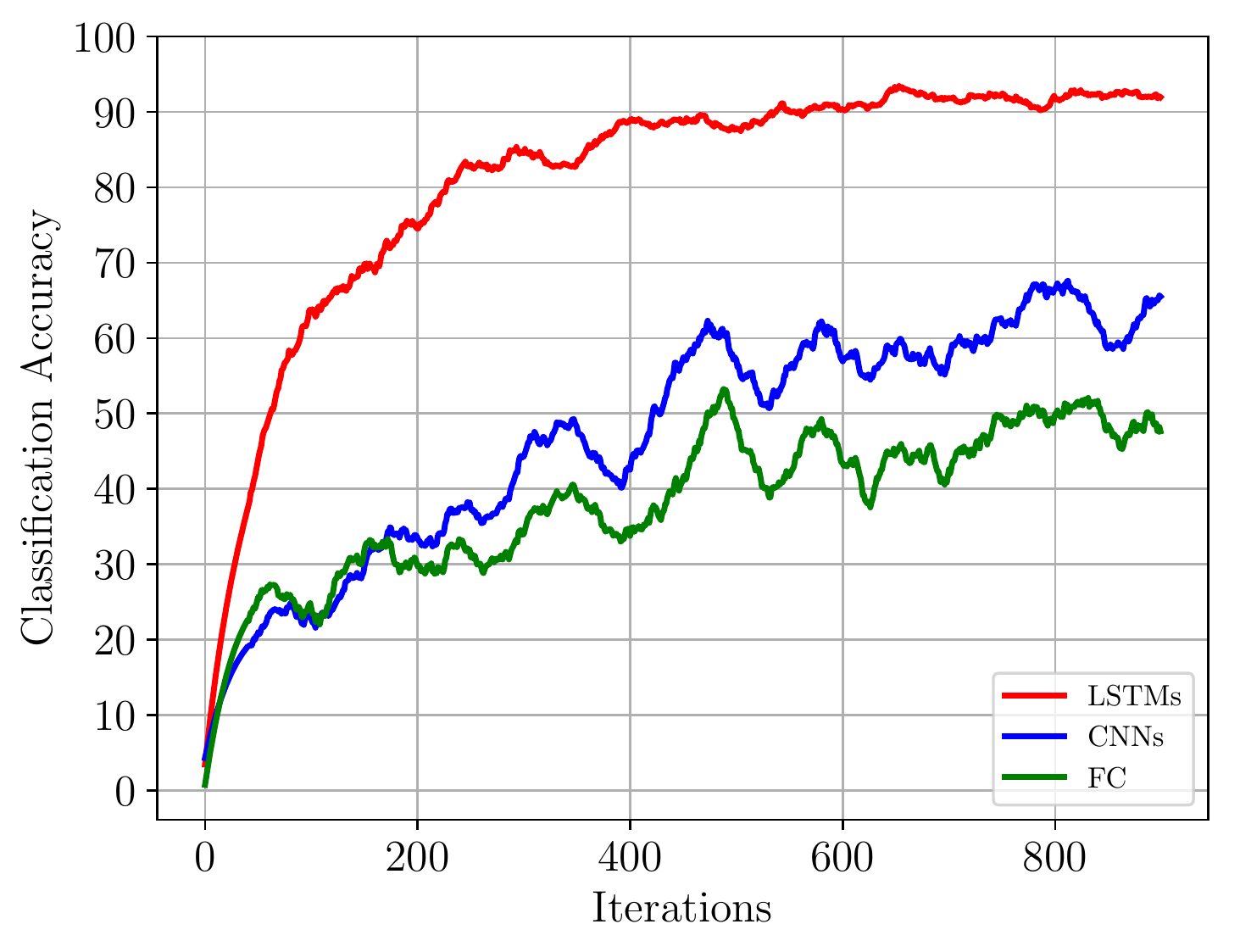}}}%
        \caption{Comparison of variation in classification accuracy during training of different ML models on dataset $1$}%
        \label{fig:accuracy_comparison}%
    \end{figure}
    \item \textbf{Average Upper Bound on Error}: 
    We calculate the average upper bound on the error in the predicted distance over all the test cases including the ones where the model does not predict the correct bin label (around $7\%$ cases as reported in the previous section). Let $x$ be the correct bin label and $y$ be the label predicted by the model. In the worst case, upper bound on the error in the distance ($E\textsubscript{max}$) for this observation will be given by
    \begin{equation} \label{eq:upper_bound_on_error}
      E\textsubscript{max} = \|x-y\|_2 \times l\textsubscript{bin} + \frac{l\textsubscript{bin}}{2}
    \end{equation}
    It is evident from the Fig.\ref{fig:error_comparison_in_cm} that the LSTM based architecture achieves the lowest error bounds. The error bounds achieved by LSTMs, CNNs and FCs are in the range of $8$cm, $25$cm and $30$cm respectively.
    We obtained an average error upper bound ($E$) of $8.67$ cm, $7.36$ cm, $8.12$ cm and $8.55$ cm on the $4$ datasets respectively using LSTM. Fig.\ref{fig:error_comparison_in_cm} depicts the variation of $E$ as the three architectures get trained on dataset $1$. The accuracies are reported on the scale of ($8.46$ m $\times$ $6.98$ m).
    \begin{equation} \label{eq:average_error_upper_bound}
        E = \frac{1}{N}\sum_{i=1}^{N} E\textsubscript{max}[i]
    \end{equation}
    where $N$ is the total number of test cases. We list all the obtained results in Table \ref{table:day_accuracy_error}.
    \begin{figure}[h!]%
        \centering
        \subfloat{{\includegraphics[width=8cm]{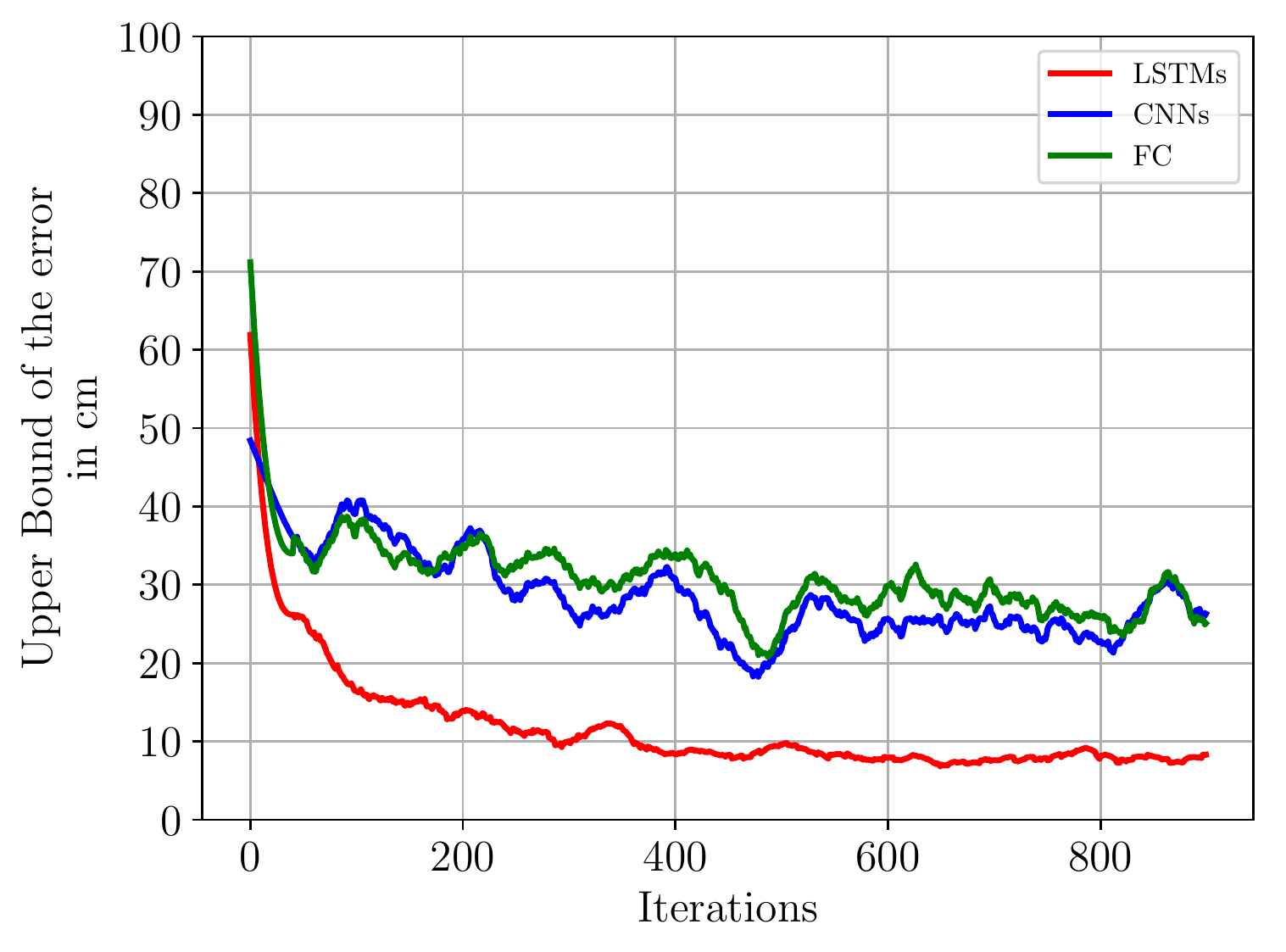}}}%
        \caption{Average Upper Bound variation as the model training proceeds on dataset $1$}%
        \label{fig:error_comparison_in_cm}%
    \end{figure}
\end{enumerate}
\begin{table}[h!]
    \centering
    \begin{tabular}{ |c|P{3cm}|P{2cm}|c| } 
        \hline
        \textbf{Dataset} & \textbf{$5.85$ cm Confidence Interval (in \%)} & \textbf{Average Upper Bound on Error} \\
        \hline
        Day 1 & $93.94$ & $8.67$ cm\\ 
        Day 2 & $92.51$ & $7.36$ cm\\ 
        Day 3 & $93.89$ & $8.12$ cm\\ 
        Day 4 & $92.99$ & $8.55$ cm\\ 
        \hline
    \end{tabular}
    \caption{Obtained accuracies over the $4$ datasets using LSTM. Our model gives an  upper bound on error in predicted distance of $5.85$ cm with confidence intervals given in column 2. Column 3 lists the average of upper bounds over all test cases}
    \label{table:day_accuracy_error}
\end{table}
Table \ref{table:accuracy_comparison_other_methods} compares the average reported error of our method with other existing works and literature. 
\begin{table}[h!]
    \centering
    \begin{tabular}{ |P{2.5cm}|P{2.2cm}|P{2.5cm}| } 
        \hline
        \textbf{Methods} & \textbf{Average Errors} & \textbf{Scale} \\
        \hline
        Ibrahim et al. \cite{CNN_Ibrahim} & $277$ cm & A City Building\\ 
        \hline
        Lukito et al. \cite{8267833} & $83$\% Classification Accuracy  & University Campus \\
        \hline
        Wang et al. \cite{WangGao} & $94$ cm  & Room of dimension $\bf{4}$ m $\times$ $\bf{7}$ m \\
        \hline
        Sadowski et al. \cite{sadowski_spachos_2018} & $48.6$ cm & Room of dimension $\bf{10.8}$ m $\times$ $\bf{7.3}$ m \\ 
        \hline
        OUR METHOD & $8.67$ cm & Room of dimension $\bf{8.46}$ m $\times$ $\bf{6.98}$ m \\
        \hline
    \end{tabular}
    \caption{Comparing average errors in indoor environments of different methods}
    \label{table:accuracy_comparison_other_methods}
\end{table}
\section{Conclusion and Future Work}
In this work, we have presented an indoor distance estimation system that uses the RSSI values provided by the commercially available standard WiFi chips. Our model successfully uses LSTMs and achieves an accuracy of $5.85$ cm with a confidence interval of around $93\%$ which is a statistically significant improvement over other approaches. Our system is self-adaptive as it can adjust its parameters to estimate the object distance in different environments accurately. We take into account the unwanted multipath fading effects without adding any extra hardware components dedicated to filtering the multipath effects selectively. Also, since this system has been developed using the existing infrastructure of WLAN, the deployment cost is very low, making it commercially viable. \par
The system predicts target object distance from the access point and thus $3$ WAP\textquotesingle s can be used to perform triangulation and thereby detect the object location. The system can be made more accurate by leveraging the fact that CSI gives more information about the channel between transmitter and receiver compared to RSSI and hence can be used to model the environment in a more sophisticated manner.
%
\addtolength{\textheight}{0cm}   
%
%
\bibliography{main}
\bibliographystyle{ieeetr}
\end{document}